\documentclass[manuscript]{aastex}

\slugcomment{To appear in Astrophysical J.}

\shorttitle{Formation scenario} \shortauthors{Jilinski et al.}

\begin{document}


\title{Dynamical evolution and spectral characteristics of
 the stellar group Mamajek~2}


\author{E. Jilinski \altaffilmark{1, 2}, V. G. Ortega\altaffilmark{1},
R. de la Reza \altaffilmark{1}, N.A. Drake \altaffilmark{1, 3}, B.
Bazzanella \altaffilmark{1}}


\altaffiltext{1}{Observat\'orio Nacional, Rua General Jos\'e
Cristino 77, S\~{a}o Cristov\~{a}o, 20921-400, Rio de Janeiro,
Brazil.}

\altaffiltext{2}{Pulkovo Observatory, Russian Academy of Sciences,
65, Pulkovo, 196140 St. Petersburg, Russia.}

\altaffiltext{3}{Sobolev Astronomical Institute, St. Petersburg
State University, Universitetskii pr. 2, Petrodvorets, 198504, St.
Petersburg, Russia.}

\email{jilinski@on.br}

\begin{abstract}

The dynamical evolution of the recently detected stellar group
Mamajek~2 is studied by means of its past 3D orbit. The past orbits
of the open clusters NGC~2516 and $\alpha$~Persei, belonging to the
so-called ``Local Association", were also computed in order to check
for a possible common past dynamical evolution of these systems. To
complete the data of the Mamajek~2 small group, we have obtained
high resolution FEROS spectra to measure the radial and also the
projected rotational velocities of its members; an estimate of its
metallicity was obtained as well. Two exceptionally low rotating
A-type stars turned out to be a strong magnetic Ap star in one case,
and a normal A0 star with near-solar metallicity in the other. The
dynamical results showed that NGC~2516 and Mamajek~2 may have had a
common origin at the age of 135 $\pm$ 5 Myr. This dynamical age
confirms the individual ages of 140 Myr for NGC~2516 and 120 $\pm$
25 Myr for Mamajek~2 obtained independently by photometric methods.
Both these groups appear to have the same solar metallicity giving
support to a common birth scenario. The dynamical approach is
showing that some bound open clusters can form in a coeval fashion
with unbound stellar groups or with associations.

\end{abstract}

\keywords{(GALAXY:) open clusters and associations: individual
\objectname{NGC~2516, Mamajek~2, $\alpha$~Persei}, stars: individual
\objectname{HD~158450, HD~160142} }

\section{Introduction}

The recent discoveries of young loose stellar groups or unbound
associations in the solar neighborhood, at distances up to about 100
pc (Zuckerman \& Song 2004, Torres et al. 2008) motivated a deeper
research of what is known as the Local Association or the Pleiades
Supercluster (Eggen 1975, 1983). Historically, this supercluster has
been conceived as a very large structure with a radius of a few
hundred parsecs containing stars with space velocities similar to
those of the Pleiades cluster. Considered like this, the Local
Association would contain different smaller structures, such as the
Pleiades, NGC~2516, $\alpha$~Persei, NGC~1039, IC~2602 open clusters
and the whole OB Sco-Cen association.

In an attempt to visualize the structural/dynamics of the Pleiades
Supercluster, Skuljan et al. (1999), based on Hipparcos astrometry,
made a 2D analysis and found among others, the existence of a
``Pleiades branch" in the $U,V$ space of velocities containing the
separated Hyades and Pleiades groups. However, Skuljan et al. (1999)
had difficulties in concluding whether this extended branch was due
either to some special feature of the potential of the Galaxy or to
the local spiral structure. Moreover, a complete mixture of ages was
in play in this branch covering some few hundred million years.

Some years ago we initiated a 3D Galactic dynamical approach to
study in detail the past evolution (ages and formation regions) of
different young moving groups related to the Sco-Cen association.
First, we studied the low-mass, loose stellar group $\beta$~Pic, the
$\epsilon$ and $\eta$~Chamaeleontis young cluster and the TW~Hya
association (Ortega et al. 2002, 2004, Jilinski et al. 2005 and de
la Reza et al. 2006). All these groups were probably originated, as
well as the younger Sco-Cen component Upper Scorpius (US), in the
mainstream of the older Sco-Cen subgroups Lower Centaurus Crux (LCC)
and Upper Centaurus Lupus (UCL) during the last 5 to 11 Myrs.

In Ortega et al. (2007) we  tackled the dynamical evolution of the
Pleiades open cluster and the AB~Dor group. Here, we investigate the
common evolution of two further members of the Local Association:
the open cluster NGC~2516, and Mamajek~2, a recent detected and
studied stellar group (Mamajek 2006). This author suggests that the
Mamajek 2 group may have formed in the same star forming region as
the clusters Pleiades, NGC~2516 and $\alpha$~Persei and the AB~Dor
group. In our previous work (Ortega et al 2007), we found that while
the Pleiades cluster and the AB~Dor group, in fact, could have had
the same origin, the $\alpha$~Persei cluster shows a completely
different past dynamical evolution. In the present work, we show
that the group Mamajek~2 and the open cluster NGC~2516 may have had
a common origin, however, again quite distinct from that of the
$\alpha$~Persei cluster.

This paper is organized as follows: In Section 2 we present the main
properties of the concerned stellar groups. Section 3 is devoted to
the presentation of the observations of the stars of the Mamajek~2
group together with the measured radial and rotational velocities.
Section 4 contains the dynamical aspects of the involved stellar
groups. Section 5 presents the spectral analysis for two low-
rotation A-type stars belonging to the Mamajek~2 group. Finally,
Section 6 is devoted to the discussion and conclusions.

\section{Main properties of the open cluster NGC~2516
         and the Mamajek~2 stellar group}

\subsection {The NGC~2516 open cluster}

NGC~2516, also called the ``southern Pleiades'' by Eggen (1972,
1983), is a rich, nearby, bright open cluster affected by small
extinction. In studying possible large gravitational tidal effects
(its present tidal radius is about 9 pc (Piskunov et al. 2008) on
this cluster, Bergond et al. (2001) noted the regular circular
geometry of its center from where two tails emerge almost
perpendicular to the Galactic plane. The total mass of NGC~2516 is
presently not well known. Values as low as 170 solar masses have
been proposed by Pandey et al. (1987) and as high as about 1000
$M_\odot$ by Dachs \& Kabus (1989). More recently a mass of $\sim$
250 $M_\odot$ for this cluster has been quoted by Piskunov et al.
(2008). In contrast, its age appears to be quite well established.
Recent literature adopts the age determined by Meynet et al. (1993)
by fitting NGC~2516 with an isochrone at 140 Myr.

Concerning the metallicity of NGC~2516, the situation does not seem
to be entirely clear. This is mainly due to the fact that bright F
stars in NGC~2516, which would normally be used to derive the
abundances, possess, in general, largely broadened spectral lines.
Terndrup et al. (2002) present a quite thorough discussion of past
and more recent metallicity determinations. Nevertheless, because of
the importance of this parameter for the present study, some
insights will be furnished. After an initial period where several
authors (see Terndrup et al. 2002) found for NGC~2516 a metallicity
of a few tenths dex below the solar, more recent analysis (Irwin et
al. 2007, Jeffries et al. 2001, Sciortino et al. 2001, Sung et al.
2002) places NGC~2516 with a metallicity close to solar. This is in
fact the conclusion of a careful spectroscopic analysis, albeit
based on only two low rotating, relatively hot stars, that gives
[Fe/H] = $+ 0.01 \pm 0.17$. Also, a photometric determination
yielded a value of [Fe/H] = $- 0.05 \pm 0.14$. As commented by these
authors, an analysis of faint G stars of NGC~2516 with larger
telescopes will be necessary to settle the question.

In our discussion we shall adopt a near solar metallicity for
NGC~2516 which then will be compared with that estimated for
Mamajek~2. In Section 4 we will see how a near solar abundance of
NGC~2516 will be indirectly compatible with its dynamical age.

\subsection {The Mamajek~2 stellar group}

This new stellar aggregate was discovered by Mamajek (2006)
based on the common parallel proper motions and similar
trigonometric parallaxes of the stars. It contains the bright B8
giant $\mu$~Ophiuchus and eight further B and A type stars. This
author proposes the coevality of this group located at a
present distance of 170 pc and with an age of $120 \pm 25$ Myr. The
scatter distance of the group members is $\pm$ 5 pc and the total
mass is of the order of 24~M$_\sun$. According to Mamajek (2006), the
half-mass radius of this cluster is 0.4 pc while its tidal radius
is of the order of 4 pc.

Adopting the canonical Initial Mass Function (IMF) of Kroupa (2001),
Mamajek (2006) proposes that the present existence of these nine
stars would imply an initial population of $\thicksim$ 200 systems.
So far the low mass members of this group have not been detected. In
this work we have  measured the radial velocities of the proposed
members in order to confirm the reality of this stellar group. In
fact, previously only two stars, ($\mu$~Oph (HD~159975) and
HD~158450) had their radial velocities measured (Mamajek 2006). No
attempt is made here to detect other cooler and low mass members.

\section{Observations of the Mamajek~2 stellar group}

\subsection{Radial velocity determinations}

High resolution spectra have been obtained for eight of the nine
known members of the Mamajek~2 group. Unfortunately, for only one
star (HD~159874) this was not possible because of bad weather
conditions. There are no data concerning the radial velocity of this
star in the literature. For these observations we used the FEROS
spectrograph attached to the 2.2 m telescope of ESO at La Silla -
Chile. The spectra were obtained using a resolution of about 48000
and covering a spectral range from 3800 to 9200 $\AA$. The main
objective of these observations was to measure the radial and
rotational projected velocities and also to determine metallic
abundances for some representative stars of the group. Standard
FEROS pipeline resources for calibration purposes have been used.

Precise radial velocities measurements for B and A-type stars are
intrinsically difficult. This is partly due to effects of their high
temperatures and large or very large rotational velocities,
especially of B stars. In addition, the few lines available for both
types of stars makes the measurements difficult. These difficulties
are reflected in the fact that there are no appropriate radial
velocities standard stars (again especially for B-type stars).
Nevertheless, the high stability of FEROS and its large spectral
dispersion, make this instrument one of the best available for
radial velocity measurements. To measure these velocities we have
followed the same methodology as previously used with hot stars of
the Sco-Cen OB association (Jilinski et al. 2006). The
cross-correlation technique, which is used for precise RV
determinations in later type stars, when applied to the hotter BA
stars can be problematical as early type stars spectra show few
absorption lines which are, in many cases, intrinsically broad (up
to a few hundreds km\,s$^{-1}$) due to stellar rotation. Because of
this we determine  the doppler shift for each measurable unblended
spectral line of He I, C II, N II, O II, Mg II, Si II and Si III, Fe
I and Fe II, Ti I and Ti II relative to their rest wavelengths. The
measured radial velocities with their error bars are listed in Table
1.

\subsection{Projected rotational velocities}

Projected rotational velocities were determined by the synthetic
spectra method, the most accurate method for $v\sin i$ determination
consisting in computing the synthetic spectrum and comparing with
the observed one (de Medeiros et al. 2006). To select the
appropriate atmospheric models, we used the $(B-V)_0$ -- $T_{\rm
eff}$ calibrations from Kenyon \& Hartmann (1995) to determine the
effective temperature. Observed values of the color indexes $B-V$
and color excesses $E_{B-V}$ were taken from Table~2 of Mamajek
(2006). For the star $\mu$~Oph we used $T_{\rm eff}=12020$~K taken
from Glagolevsky (1994). A value of the surface gravity equals to
$\log g = 4.0$ was adopted for the synthetic spectra calculations.

In general, in order to determine $v\sin i$, we selected two
spectral regions: 7770 - 7780~\AA\ containing the O\,{\sc i}
infrared triplet and 4470 - 4490~\AA\ containing the Mg\,{\sc ii}
4481.2 line. The values of $v\sin i$ obtained for the Mamajek 2
group are presented in Table~1. Six stars of the group have high
rotation compatible with their spectral types (Royer et al. 2002,
2004, Abt \& Morrell 1995) and two stars, HD~158450 and HD~160142,
show low projected rotation velocities. Star HD~158450 is a peculiar
A-type star with an important magnetic field (Kudryavtsev et al.
2006) believed to be responsible for its low rotation. Star
HD~160142, is a normal A0 star, probably seen pole-on. For
HD~160142, besides the above mentioned spectral lines, we consider
also the spectral synthesis of the  Fe\,{\sc ii} lines at 4489.183,
4491.405, and 6147.741~\AA\ as well as the Ti\,{\sc ii} lines at
4464.449, 4468.507, and 4488.325~\AA, which allowed us to determine
the projected rotation velocity of this star with an even higher
precision. In the case of the chemically peculiar star HD~158450
which has spectral lines broadened by the magnetic field effects, we
performed an analysis of the magnetically splitted Fe\,{\sc ii} line
at 6149.258~\AA.

\section{Dynamical evolution calculations}

To study the dynamical evolution of stellar groups it is necessary
to integrate back in time the 3D orbits of the stars. Beginning with
the present distance or initial $XYZ$ positions relative to the Sun
and with the presently observed spatial velocities $(UVW)$, the 3D
past orbits are calculated using a modeled Galactic potential. More
details of the adopted methodology can be found in previous works
(Ortega et al. 2002, 2004, 2007, Jilinski et al. 2005 and de la Reza
et al. 2006).

For small stellar groups and stellar associations, with ages less
than $\thicksim$ 12 Myr, it was possible to find the dynamical ages
and the respective birthplaces by determining the first maximum
confinement of the individual orbits. This was the case for the
$\beta$~Pic and TW~Hya associations (Ortega et al. 2002, 2004 and de
la Reza et al. 2006). For clusters, instead, it is advantageous to
work with the mean values of the velocities of the stars in order to
minimize input errors, especially in the case of a relatively large
age of the system. In this case, we evolve the group with unknown
age together with another system, preferentially a cluster, having a
reasonably well determined age. If both stellar systems attain a
maximum approach at some time and if such approach occurs with low
relative velocity, we consider this time as the age of the first
group. This technique was employed by us in a study of the Pleiades
open cluster and the AB~Dor association (Ortega et al. 2007). We
shall use this methodology in the present work.

Dynamical calculations can give good results only if present
positions and space velocities of cluster members have the most
precise values as possible. For this reason stars with Hipparcos
(ESA, 1997) astrometric data are preferentially used in this work.
Likewise, radial velocities of high quality are important. For the
Mamajek~2 group we used the mean value of the $v_{\rm rad}$ of the
Hipparcos stellar members which were measured in this work. These
measurements are shown in Table 1. Concerning the mean radial
velocities for NGC~ 2516, we considered the following data: $23.8
\pm 0.3$ km\,s$^{-1}$ obtained for 24 members by Jeffries et al.
(1998); $22.0 \pm 0.2$ km\,s$^{-1}$ for 22 members (Gonz\'alez \&
Lapasset 2000); $22.7 \pm 0.4$ km\,s$^{-1}$ for 14 members (Robichon
et al. 1999) and finally $24.2 \pm 0.2$ km\,s$^{-1}$ for 57 members
by Terndrup et al. (2002). We also check the possibility of common
evolution of the Mamajek~2 group with the $\alpha$~Persei cluster,
as suggested by Mamajek (2006). For the $\alpha$~Persei cluster we
used the published $UVW$ values of Robichon et al. (1999) obtained
using 46 Hipparcos star members. Spatial velocities given by Makarov
(2007) were also considered for this cluster, noting however, that
in this case the distances were kinematically determined.

For NGC~2516, the use of a set of different mean radial velocities
resulted in different past orbits, clearly indicating that the
dynamical calculations are sensitive to this quantity, especially
for ages larger than 100 Myr. The sample of Terndrup et al (2002)
contains a largest number of stars. The mean radial velocity quoted
by these authors were obtained combining their own observed
velocities (33 determinations) with those in Jefferies et al. (1998)
for six common members. This also allowed to have an estimate of the
systematical error. The best result indicating a very probable
common origin of NGC~2516 and Mamajek~2 groups is that obtained
using the data of Terndrup et al. (2002). This can be appreciated in
Figure~1 where the past 3D distance between these two groups is
shown as a function of time in the past. A maximum approach or a
minimum distance of nearly 20 pc is obtained at $-135 \pm 5$ Myr.
The uncertainty in the age was estimated through Monte Carlo
simulations using 1000 realizations. In Figure~1 (low panel) we show
the past evolution of their relative velocity and angle of
approximation.

In Figure~2 we show a similar analysis for the past evolution of the
$\alpha $~Persei cluster and the Mamajek~2 stellar group. It shows
that, for both cases Robichon and Makarov, these systems do not
achieve any approach in their evolution. Dynamically there is no
relation between them.

Figure 3 shows the mean 3D orbits of Mamajek~2 and the cluster
NGC~2516 projected on to the Galactic plane (XY) and on the plane
(YZ)perpendicular to the Galactic plane.

\section{The magnetic chemically peculiar star HD~158450}

One of the stars of Mamajek 2 group, HD~158450, shows a highly
peculiar spectrum. This star was included in the list of ``the
brighter stars of astrophysical interest in the southern sky'' by
Bidelman \& MacConnell (1973) based on the Michigan blue
objective-prism survey of the southern sky as a peculiar A star of
the Sr-Cr-Eu type. The presence of a magnetic field on the surface
of this star was recently discovered by Kudryavtsev et al. (2006)
from spectropolarimetric observations of a sample of chemically
peculiar stars at the 6-m telescope of the SAO RAS, Russia.

We have only one high resolution spectrum of HD~158450 obtained at
June 2, 2007 (MJD = 54253.31316).  Analysis of this spectrum
indicated the presence of a strong magnetic field resulting in the
magnetic splitting of some spectral lines. The most prominent
spectral feature is the Fe\,{\sc ii} 6149.258~\AA\ line, commonly
used for magnetic field strength determination, due to the specific
Zeeman pattern of this line consisting of two $\pi$- and two
$\sigma$-components with the same wavelength shift (Mathys et al.
1997). In unpolarized light the profile of this line in the presence
of a magnetic field is a simple doublet (see Figure 4). The
wavelength shift between the red and blue components of the Fe\,{\sc
ii} 6149.258~\AA\ line in our spectrum of HD~158450 is
$\Delta\lambda_Z = \lambda_{\rm r} - \lambda_{\rm b} = 0.530\pm
0.010$~\AA.The mean magnetic field modulus (the line-intensity
weighted average over the visible stellar hemisphere of the modulus
of the magnetic vector) can be estimated by the equation (St\"utz et
al. 2003):

$$\frac{\Delta\lambda_Z}{2} = 4.67\cdot 10^{-13}\cdot g_{\rm eff}\cdot \lambda^2 \cdot \langle H \rangle,$$

\noindent where  $\Delta\lambda_Z$ - measured Zeeman splitting in
\AA, $g_{\rm eff} = 1.35$ - effective Land\'e factor, $\lambda$ -
central wavelength of the unshifted line in \AA, and $\langle H
\rangle$ - mean magnetic field modulus in Gauss. Using this formula
we find the mean magnetic field modulus $\langle H \rangle=11100\pm
200$~G.

In Kudryavtsev et al. (2006) the mean longitudinal component of the
magnetic field was determined to be ${\langle {B_l}^2
\rangle}^{\frac{1}{2}} = 1570\pm 180$~G, whereas the individual
values of the longitudinal magnetic field vary for different dates
from $-2920 \pm 200$~G to $+810\pm 240$~G, indicating strong
variation of the magnetic field strength with stellar rotation.

The high resolution spectrum obtained by us permitted to measure the
projected rotational velocity of HD~158450 with high precision.
Kudryavtsev et al. (2007) determined $v\sin i = 20\pm
2$~km\,s$^{-1}$, a value close to the lower limit for a projected
rotational velocity of 18~km\,s$^{-1}$ which can be achieved with
moderate-resolution spectra ($R=15\,000$) used in the above
mentioned paper. Kudryavtsev et al. (2007) explicitly say in their
article that rotational velocities of magnetic stars have to be
determined using high resolution spectra and by  the comparison of
observed spectra with the synthetic ones. We note that the
rotational velocities in Kudryavtsev et al. were estimated by FWHM
measurements of two Fe\,{\sc ii} lines (4508 and 4491 \AA) having
low Lande factors.

The approximation of the Fe\,{\sc ii} 6149.258 \AA\ magnetically
split line by a synthetic spectrum showed that this star has the
significantly lower projected rotational velocity of $v\sin i = 9\pm
1$~km\,s$^{-1}$.

Our determination of the radial velocity of this star ($v_{\rm
rad}=-17.3$~km\,s$^{-1}$) is in perfect agreement with the value of
$v_{\rm rad}=-17.2\pm 1.4$~km\,s$^{-1}$ obtained by Kudryavtsev et
al. (2007). Previously, Grenier et al. (1999) found the value of
$-22.0\pm 4.2$~km\,s$^{-1}$ for the radial velocity of this star.
Although the difference in radial velocity is rather large, we note
that Grenier at al. (1999) measured the radial velocity of this star
by correlation with template of the same spectral class. The
peculiar nature of the spectrum of HD~158450 could lead to some
discrepancy. That is why the double, or multiple nature of this
star, as pointed out in the SIMBAD database, needs further
investigation by monitoring its radial velocity.

We detected resolved magnetically split lines in this this
chemically peculiar star. However, a careful determination of the
atmospheric parameters and a detailed abundance analysis of this
star is beyond the scope of this paper. We emphasize that HD~158450
is a member of the Mamajek 2 stellar group which has a quite well
determined age. A detailed study of this star would then be very
important for the understanding of the origin and evolution of
stellar magnetic fields, a problem not clarified until now.

\section{A chemical analysis of HD~160142}

The star HD~158450 being an Ap star, is not appropriate for the
metallicity estimation of the Mamajek~2 group. We invested then in
an analysis of the normal low rotating A0 star HD~160142 using the
last version of the well-known MOOG program (Sneden 1973). In order
to determine the atmospheric parameters of HD~160142, we measured
the equivalent widths of neutral and ionized iron lines whose
oscillator strength values ($\log gf$) were analyzed by Lambert et
al. (1996). Our analysis was somewhat hampered by the relatively
noisy observed spectrum of this star what makes difficult the
measurement of faint Fe\,{\sc i} lines with equivalent widths lower
than 10 - 15~m\AA. After eliminating suspected blends and very weak
lines, whose equivalent widths could not be measured in our spectrum
with high precision, we used 25 Fe\,{\sc i} and 17 Fe\,{\sc
ii} lines. 
Following the usual iterative procedure, we derived the effective
temperature and microturbulent velocity by requiring the iron
abundance to be independent of the excitation potential and of the
equivalent width (Fig. 5). The surface gravity was derived from the
ionization equilibrium by finding the value for which the iron
abundances from Fe\,{\sc i} and Fe\,{\sc ii} coincide. The Kurucz's
(1993) grid of atmospheric models was used in the calculations. The
following atmospheric parameters (effective temperature, surface
gravity, and microturbulent velocity) were derived: $T_{\rm
eff}=9320$~K, $\log g = 3.8$, and $\xi_{\rm m}=2.07$~km\,s$^{-1}$.
In this case, the metallicity of HD~160142 is $\log\varepsilon({\rm
Fe}) =7.62\pm 0.05$, i.e. [Fe/H]$ = +0.10$. We must note however,
that even if the $T_{\rm eff}$ value coincides with the photometric
temperature based on the $T_{\rm eff}$ {\it versus} $(B-V)_0$
calibration (Kenyon \& Hartman 1995), the obtained value of the
surface gravity ($\log g = 3.8$) is too low for a star of such a
temperature and age.

A different method was then explored to estimate the surface gravity
of HD~160142 by analyzing its position on the HR diagram. Knowing
the effective temperature $T_{\rm eff} =9320$~K from the photometric
data, the surface gravity can be inferred from internal structure
models. Using isochrones from the models of Lejeune \& Schaerer
(2001) for solar metallicity ($Z=0.02$) and assuming for HD~160142
an age of $\log t = 8.10$ or 126 Myr (similar to the age of the
Mamajek~2 group) for HD~160142 we obtained for the surface gravity
the value $\log g = 4.27$ resulting in a solar iron abundance for
the Fe\,{\sc i} lines ($\log\varepsilon({\rm Fe}) = 7.51\pm 0.05$)
and a higher abundance for the Fe\,{\sc ii} lines
($\log\varepsilon({\rm Fe}) = 7.69\pm 0.08$).

This discrepancy between the iron abundances derived from the
Fe\,{\sc i} and Fe\,{\sc ii} lines may be due to different reasons.
Problems with the Fe\,{\sc i}/Fe\,{\sc ii} ionization balance have
been reported for a wide range of stars (e.g. Allende Prieto et al.
1999). Recently, Yoon et al. (2008) analyzing a high-resolution
spectrum of the known A-type star Vega have demonstrated the effects
of rotation on the derived abundances. They found that, in the case
of Vega, the rotation induces an iron ionization imbalance amounting
to $\sim 0.35$~dex, but in the opposite sense of that induced by
departures from LTE. Thus, if HD~160142 is a high rotating star seen
nearly pole-on, we have to use the model with a lower value of the
surface gravity forcing to achieve the ionization balance, i.e.
equality between the iron abundance derived from Fe\,{\sc i} lines
and that derived from Fe\,{\sc ii} lines. We also note, that in the
case of HD~160142 the microturbulent velocity
($\xi=2.07$~km\,s$^{-1}$) was determined using the relatively weak
Fe\,{\sc i} lines, and may be different for the more intense
Fe\,{\sc ii} lines. On the other hand, noise can perturb somehow a
proper determination of the continuum introducing an error in the
equivalent width measurements of the weak Fe\,{\sc i} lines.

Error estimates of the derived atmospheric parameters are not
straightforward. An uncertainty of $\pm 0.02$~mag in the correction
for interstellar reddening of the observed $(B-V)$ index results in
an uncertainty of about $\pm 200$~K in the ``photometric''
temperature. The typical uncertainty in the microturbulent velocity,
obtained in the usual way by finding the value which provides iron
abundance independent of the equivalent width of the Fe\,{\sc i}
lines, is about $\pm 0.2$~km\,s$^{-1}$. Taking into account
uncertainties in the atmospheric parameters, equivalents widths
determinations, non-LTE effects, and possible effects of rotation
(if HD~160142 is indeed highly rotating A star seen nearly pole-on)
we estimate that HD~160142 has near solar metallicity
$\log\varepsilon({\rm Fe})=7.62\pm 0.10$. Nevertheless, taking into
consideration all possible sources of uncertainties, we have to note
that the real value of the error can be somewhat greater than 0.10
dex.

It is then fair to conclude that until other cooler stars of Mamajek
2 will be discovered and analyzed, a solar abundance for the A0-type
star HD~160142 may be adopted as representative of the metallicity
of this stellar group.


\section {Discussion and conclusions}

Calculations of the dynamical history of young stellar groups is a
new source of astrophysical knowledge. In the present work we have
investigated the past evolution of three stellar groups, belonging
to what historically has been known as the Local Association or the
Pleiades Supercluster, with the aim to disentangle the dynamical
evolution of their components and make evident possible relations
existing between them. These three groups are the open clusters
NGC~2516 and $\alpha$~Persei and the recent detected small group
called Mamajek~2 (Mamajek 2006).

We have obtained new data new data for eight of the nine known
members of the Mamajek 2 group by means of high resolution FEROS
spectra. This allows to measure the radial velocities with a method
especially devised for hot stars of spectral types B and A (Jilinski
et al. 2006)(see section 3.1). Taking advantage of the high quality
of the data, rotational, projected velocities as well as information
concerning the nature and chemical abundances of some stars have
been obtained. Among the A-type stars of Mamajek~2 we have found
only two systems with low values of $v\sin i$. One of them,
HD~158450 turned out to be a magnetically Ap star while the other,
HD~160142, is a normal A0 star, probably observed in a pole-on
orientation. The mean magnetic field modulus of HD~158450 measured
by us is $\langle H \rangle=11100$~G, confirming the strong field
previously found for this star by Kudryavtsev et al. (2006) for this
star. However it is important to note that knowing the value of the
magnetic field of a star of known age can be an important ingredient
to solve the puzzle of the origin and evolution of stellar magnetic
fields. The chemical analysis of the normal A-type star HD~160142
belonging to the Mamajek~2 group, indicates that it has a near solar
metallicity.

The new measured radial velocities of Mamajek~2 members were used as
input to calculate the $UVW$ components of the spatial velocity of
each star and with them the mean 3D past orbit of the group. All
these stars have Hipparcos astrometry. The dynamical calculations
produced the following results: the open cluster NGC~2516 and the
Mamajek~2 group approached to a minimal distance of about 20 pc at
$-135 \pm 5$ Myr. This approach took place in such a way that the
angle between the velocity vectors of these two stellar systems was
of only 20 degrees and their relative velocity was fairly low ($\sim
$ 12 km s$^{-1}$). Taken together, all these factors point to a
probable common origin of these two structures. We note that the
obtained dynamical age of $-135\pm 5$ Myr for both groups is
consistent with the ages found for them in the literature, and
obtained independently using photometric isochrones. Furthermore,
both groups appear to have similar solar metallicities, supporting a
common birth scenario. Another more indirect confirmation, at least
for NGC~2516, is the fact that if the metallicity of NGC~2516 was
lower then the solar one as it was suggested before by some authors,
the photometric age of this cluster would be larger, of the order of
180 Myr (see Terndrup et al. 2002). A solar metallicity of NGC~2516,
presently more accepted, supports then the dynamical age found in
this work.

Contrary to the case of NGC~2516, the past orbit of the star cluster
$\alpha$ Persei does not show any indication of a past approach with
the Mamajek~2 group (Figure 2).

On the basis of age and kinematics similarities, Mamajek (2006)
suggested that the group Mamajek~2, the open clusters Pleiades and
$\alpha$~Persei and the AB~Dor association could have formed in the
same  star-forming complex. Our dynamical studies are showing a
different scenario in which the AB~Dor group can be considered to be
associated with the Pleiades cluster (Ortega et al. 2007), whereas
the Mamajek~2 group is associated with the open cluster NGC~2516 but
not with Pleiades nor with the $\alpha$~Persei cluster. So the
dynamical studies carried out by us point to common formation of a
bound structure and a small or large unbound star aggregate. Such a
scenario has some resemblance with the model proposed by Kroupa et
al. (2001). This model is based on N-body numerical calculations
designed to study the formation of open clusters despite the
expected gas expulsion resulting from the presence of O type stars.
Starting from a configuration of the Orion Nebular Cluster type, the
resulting one resembles the Pleiades open cluster. According to the
model the cluster so formed should be surrounded by an expanding
stellar group of the same age as the cluster.

\acknowledgments

We thank the referee for the valuable remarks and comments that
helped to improve this paper. EGJ thanks FAPERJ for the financial
support under the contract E-26/153.045/2006. We also thank Dr.
Licio da Silva who realized the spectral observations.

\begin{table}
\begin{center}
\caption{Measured radial and rotational velocities [km $\cdot$
s$^{-1}$] of the stars members of the Mamajek~2 stellar group}

\begin{tabular}{ccrrc}
\tableline\tableline
Star     &  ${V_{\rm rad}}$ & $v\sin i$ &  MJD 54250 + \\
\tableline
HD~158450 & $-17.3 \pm 1.0 $  & $9 \pm 1 $  & 3.31316537  \\
HD~158838 & $-19.0 \pm 2.0 $ & $197\pm 5$  & 3.29759192   \\
HD~158875 & $-19.7 \pm 1.0 $ & $110\pm 5$  & 3.35234547   \\
HD~159209 & $-18.4 \pm 1.0 $ & $137\pm 5$  & 3.33641668   \\
HD~159975 & $-19.6 \pm 3.0 $ & $210\pm 10$ & 2.40322284   \\
HD~160037 & $-19.5 \pm 3.0 $ & $200\pm 5$  & 2.40675644   \\
HD~160038 & $-18.8 \pm 3.0 $ & $300\pm 10$ & 2.39033997   \\
HD~160142 & $-22.8 \pm 1.0  $ & $8.5\pm 0.5$ & 2.37835364 \\
\tableline
\end{tabular}
\end{center}
\end{table}



\begin{table}
\begin{center}
\caption{Kinematics, age and metallicity of NGC~2516 cluster and
Mamajek~2 stellar group}

\begin{tabular}{lccllclll}
\tableline\tableline
Property
      & Units &   NGC~2516          &  &  & Mamajek~2           &  &   \\
\tableline
X, Y, Z &[pc]                      & $22.2, -332.1, -94.6$ &  &  & $~164.6, ~48.9,   ~39.3$ &  &   \\
U, V, W & [km $\cdot$ s$^{-1}$]    & $-17.3, -25.2 -4.1, $  &  &  & $-12.5, -24.1, -4.9$  &  &   \\
Age     & [Myr]                    &  $140$                &  &  &    $  135 \pm 5$    &  &   \\
Metallicity [Fe/H]      &    dex      &  $+0.01 \pm 0.07 $    &  &  &    $ +0.1 \pm 0.2$  &  &   \\
\tablenotetext{} {X, Y, Z - Heliocentric coordinates.}
\tablenotetext{} {U, V, W - Heliocentric velocity components. For
NGC~2516 U, V, W are calculated on the basis of Terndrup et al.
(2002) radial velocity determinations. For Mamajek~2 group U, V, W
are calculated using distance and proper motions from Mamajek (2006)
and the radial velocity newly measured by us.} \tablenotetext{}{For
NGC~2516 the value of the metallicity is from Terndrup et al.
(2002).}
\end{tabular}

\end{center}

\end{table}

\begin{figure}
\epsscale{1.5} \plottwo {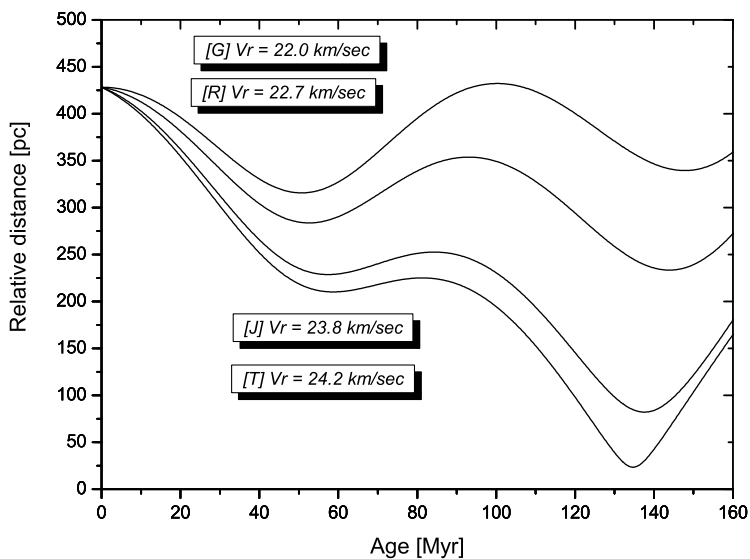}{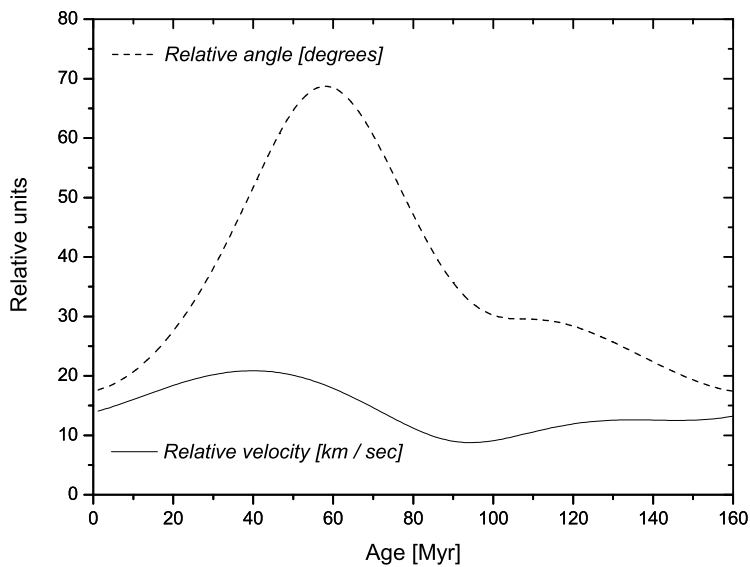} \caption{The upper panel
shows the relative distance between Mamajek 2 group and the cluster
NGC 2516 as a function of time for Gonzalez [G], Robichon [R],
Jeffries [J] and Terndrup [T] initial data sets. The lower panel
presents the time evolution of the relative velocity and the angle
between the velocity vectors of NGC~2516, using the data from
Terndrup et al. (2002), and of the Mamajek 2 group.\label{fig2}}
\end{figure}

\begin{figure}
\epsscale{1.5} \plottwo {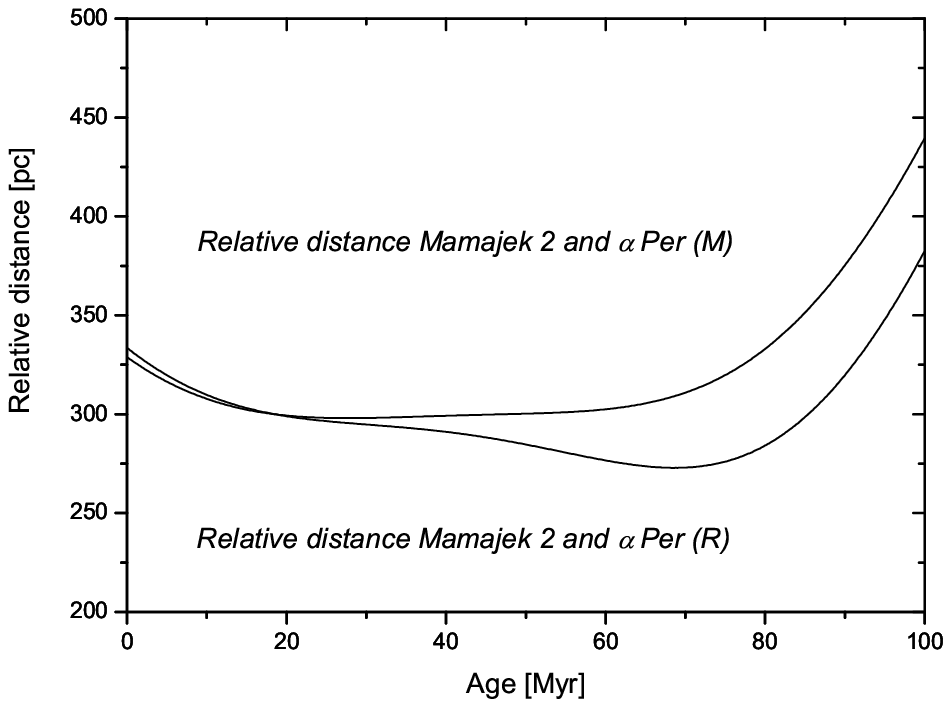}{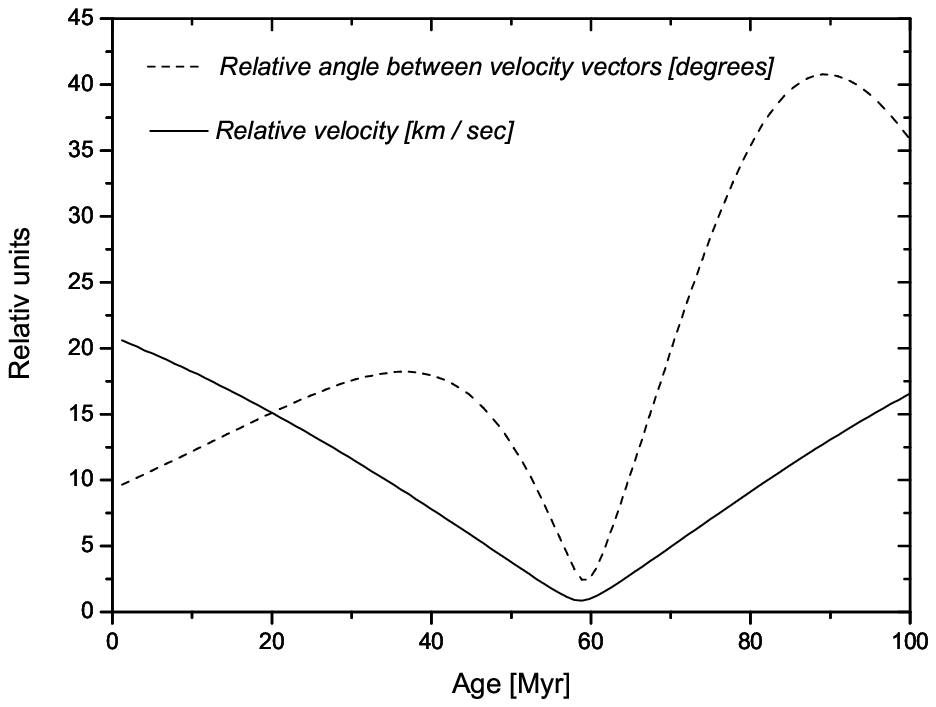} \caption{The upper panel
shows the relative distance between Mamajek 2 group and $\alpha$
Persei cluster as function of time for Makarov (2007) and Robichon
(1999) initial data sets (M \& R respectively). The lower panel
presents the time evolution of the relative velocity and the angle
between the velocity vectors of Mamajek 2 group and $\alpha$ Persei
cluster.\label{fig3}}
\end{figure}

\begin{figure}
\epsscale{1.5} \plottwo {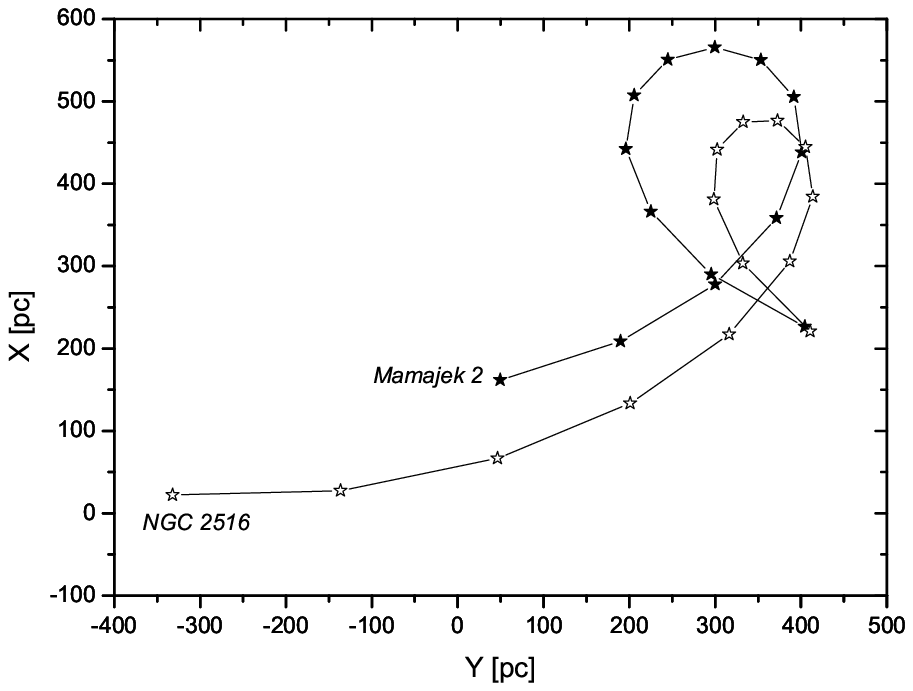}{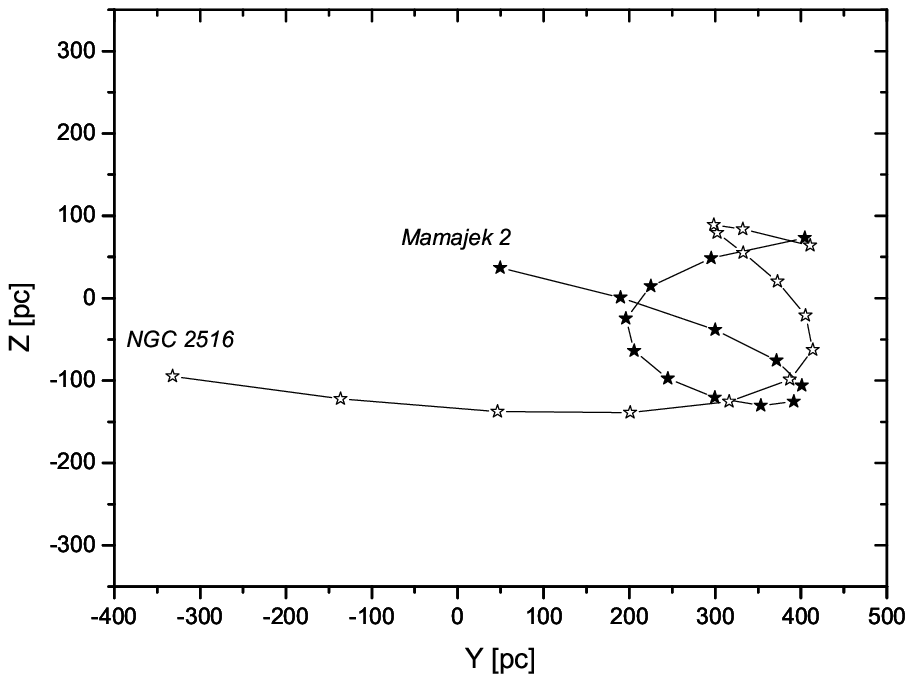} \caption{Past orbits of
NGC~2516 cluster and Mamajek 2 group. The orbit of NGC~2516 was
calculated with radial velocities from Terndrup et al. (2002). Each
interval between the symbols corresponds to 10 Myr time interval.
The first (left) points on the orbits correspond to the current
situation and the last (right) ones to the age of -135 Myr. The
upper panel shows the orbits projected on the Galactic equatorial
plane (XY) and the lower panel on the perpendicular (YZ) plane.
\label{fig6}}
\end{figure}

\begin{figure}
\epsscale{.80} \plotone{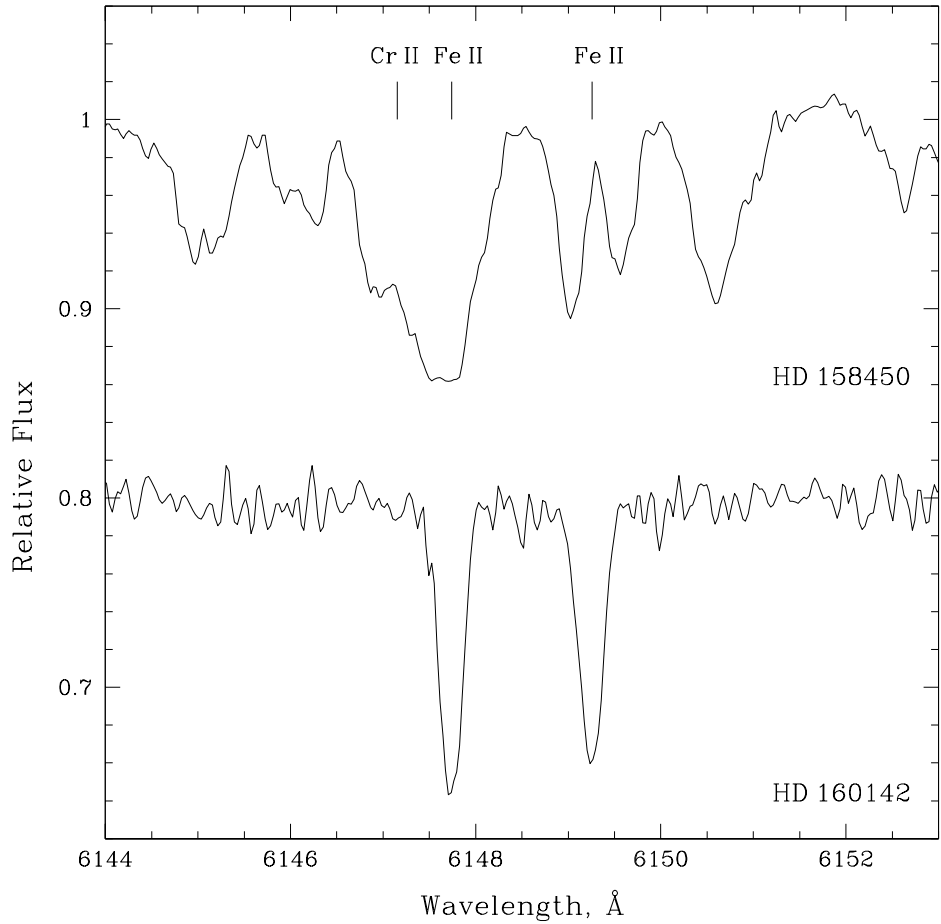} \caption{ Spectrum of the Ap star
HD~158450 showing the lines Cr\,{\sc ii} 6147.154~\AA, Fe\,{\sc ii}
6147.741~\AA, and Fe\,{\sc ii} 6149.258~\AA\ ({MJD = 54253.31316}).
Note the direct magnetic splitting of the Fe\,{\sc ii} 6149.258~\AA\
line. The spectrum of the normal low-rotating A star HD~160142
{(shifted  in intensity by 0.2)} is shown below { for comparison}.
\label{fig4}}
\end{figure}

\begin{figure}
\epsscale{.80} \plotone{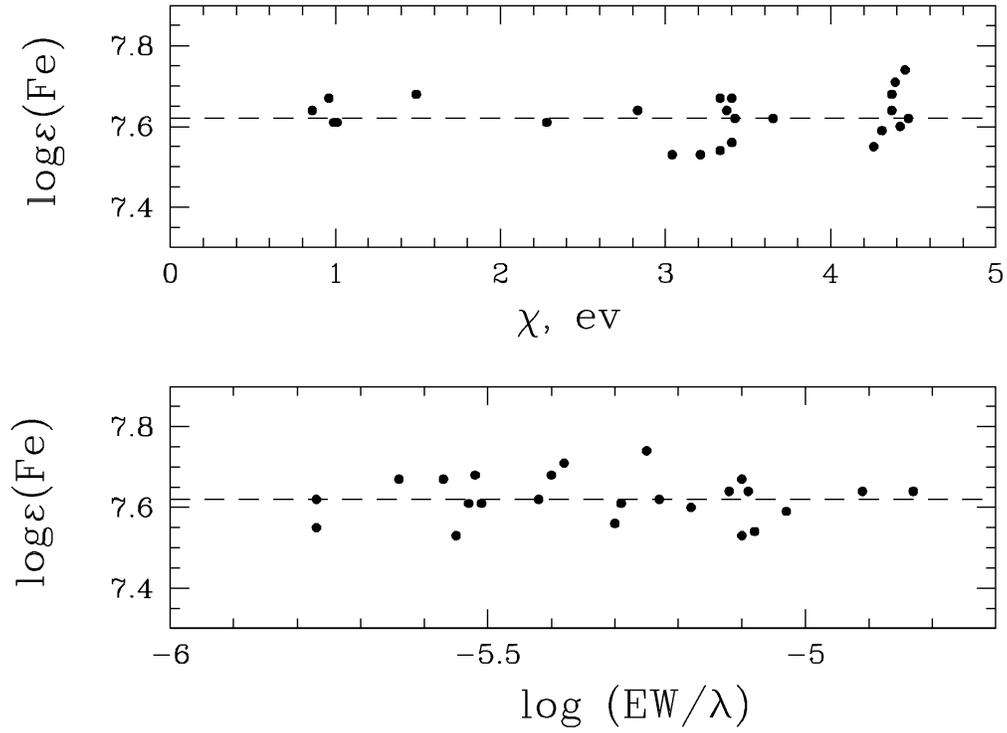} \caption{HD~160142. {\it Top:} Iron
abundances from Fe\,{\sc i} lines vs. excitation potential. {\it
Bottom:} Iron abundances from Fe\,{\sc i} lines vs. {reduced
equivalent width. The calculations were done with the adopted
atmospheric model $T_{\rm eff} = 9320$~K, $\log g = 3.8$, and
$\xi_{\rm m} = 2.07$~km\,s$^{-1}$.}
\label{fig5}}
\end{figure}

\end{document}